\documentclass[preprint,12pt]{elsarticle}

\usepackage{amssymb}

\journal{Computerized Medical Imaging and Graphics}

\begin{document}

\begin{frontmatter}

\title{Anatomically Consistent Segmentation of Organs at Risk in MRI with Convolutional Neural Networks}

\author[a]{Pawel~Mlynarski }
\author[a]{Herv\'e~Delingette}
\author[b]{Hamza Alghamdi}
\author[b]{Pierre-Yves~Bondiau}
\author[a]{Nicholas~Ayache}
\address[a]{Universit\'e C\^ote d'Azur, Inria, Epione research team, France.}
\address[b]{Universit\'e C\^ote d'Azur, Centre Antoine Lacassagne, France.}

\begin{abstract}
Planning of radiotherapy involves accurate segmentation of a large number of organs at risk, i.e. organs for which irradiation doses should be minimized to avoid important side effects of the therapy. We propose a deep learning method for segmentation of organs at risk inside the brain region, from Magnetic Resonance (MR) images. Our system performs segmentation of eight structures: eye, lens, optic nerve, optic chiasm, pituitary gland, hippocampus, brainstem and brain. We propose an efficient algorithm to train neural networks for an end-to-end segmentation of multiple and non-exclusive classes, addressing problems related to computational costs and missing ground truth segmentations for a subset of classes. We enforce anatomical consistency of the result in a postprocessing step, in particular we introduce a graph-based algorithm for segmentation of the optic nerves, enforcing the connectivity between the eyes and the optic chiasm. We report cross-validated quantitative results on a database of 44 contrast-enhanced T1-weighted MRIs with provided segmentations of the considered organs at risk, which were originally used for radiotherapy planning. In addition, the segmentations produced by our model on an independent test set of 50 MRIs are evaluated by an experienced radiotherapist in order to qualitatively assess their accuracy. The mean distances between produced segmentations and the ground truth ranged from 0.1 mm to 0.7 mm across different organs. A vast majority (96 \%) of the produced segmentations were found acceptable for radiotherapy planning.
\end{abstract}


\end{frontmatter}

\section{Introduction and related work}
\label{section_intro}
Malignant tumors of the central nervous system cause more than 200 000 deaths per year worldwide \cite{vos2016global}. Many brain cancers are treated with radiotherapy, often combined with other types of treatment, in particular surgery and chemotherapy. Radiotherapy planning requires segmentation of target volumes (visible tumor mass and areas likely to contain tumor cells) and anatomical structures surrounding lesions. The segmented volumes are used for computation of optimal irradiation doses, with the objective of maximizng irradiation of cancer cells while minimizing damage of neighboring healthy structures, called \textit{organs at risk} (OAR). Magnetic Resonance (MR) images \cite{bauer2013survey} are commonly used for imaging of tumors and organs in the brain region. In this paper, we address the challenging problem of multiclass segmentation of organs in MRI of the brain.\newline

\begin{figure}[b!]
\centering
\includegraphics[width=1.0\textwidth]{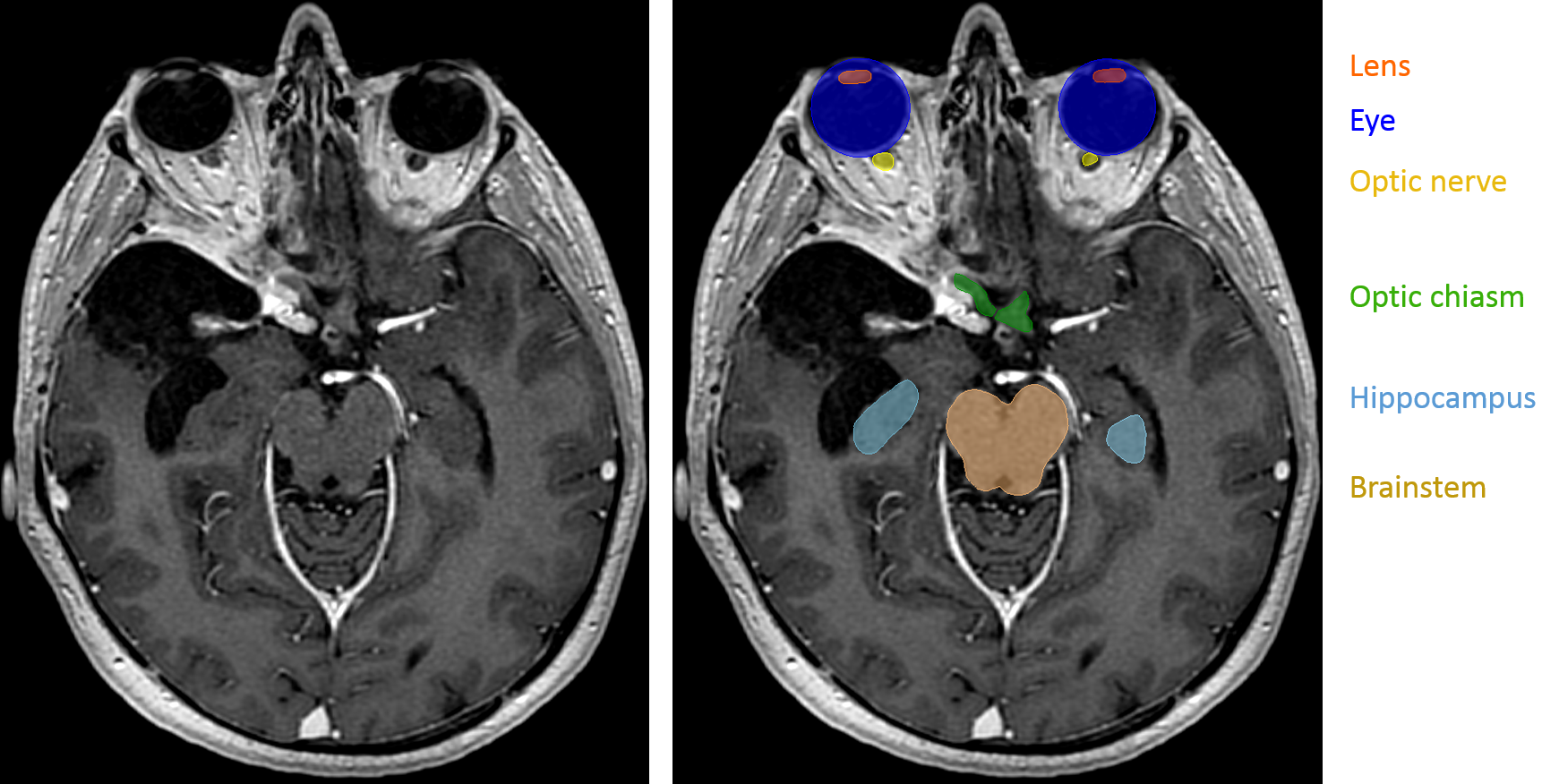}
\caption{Segmentation of organs at risk in radiotherapy planning. Left: T1-weighted MRI acquired after injection of a gadolinium-based contrast agent. Right: manual annotations of several organs at risk. In contrast to standard segmentations problems, one voxel may belong to zero or several classes (for instance, the eye and the lens).}
\label{fig_intro}
\end{figure}

Delineation of organs at risk is today manually performed by experienced clinicians. Due to a large number of structures to be accurately segmented, the segmentation process takes usually several hours per patient. Manual segmentation represents therefore a very high cost and eventually delays the beginning of the therapy. Moreover, a high intra-observer and inter-observer variability is observed \cite{brouwer20123d}. Automatic methods for segmentation of organs at risk are therefore of particular interest. We can distinguish two main types of approaches proposed in the literature.\newline\newline
The first type of methods corresponds to atlas-based approaches \cite{ciardo2017atlas,bondiau2005atlas,alchatzidis2015local}. The input image is typically registered to one \cite{commowick2008atlas,commowick2009using} or several \cite{ramus2010assessing,ramus2010multi} annotated images, from which the segmentation is extrapolated. When multiple atlases are used, the candidate segmentations may be combined, for instance, by voting strategies \cite{ramus2010multi} or by the STAPLE algorithm \cite{warfield2004simultaneous}. An important advantage of atlas-based methods is to produce anatomically consistent results. However, their main drawback is their limited generalization capacity. The important variability between cases results not only from the natural anatomical differences between patients but also from pathological factors. In particular, healthy organs are deformed by growing tumors, which may appear at different locations and which are typically not present in atlases. Some organs may even be missing because of surgeries undergone previously by the patient.
\newline
The second group of approaches is based on a discriminative classification of voxels with machine learning models such as Random Forests \cite{criminisi2010regression,criminisi2013regression,gauriau2015multi} or Convolutional Neural Networks (CNN) \cite{lecun1995convolutional}. These discriminative methods are less constrained than atlas-based approaches and may therefore better adapt to the diversity of cases. However, in general, voxelwise classifiers may produce results which are inconsistent in terms of shapes and locations of organs.\newline\newline
Organs at risk in the brain region have complex shapes and are surrounded by other structures sharing similar voxel intensities in MRI. Moreover, there are large differences related to acquisition of MRI, especially when images come from different medical centers. In order to segment organs from MRI, a complex and abstract information has therefore to be extracted. Convolutional Neural Networks are suitable for this task, as they have the ability to automatically learn complex and relevant image features. In this paper, we propose a system based on CNNs for multiclass segmentation of organs at risk in brain MRI. Anatomical consistency of the result is enforced in a postprocessing step.\newline\newline
In this paper we assume non-exclusive classes, i.e. that one voxel may belong to zero or several classes (Fig. \ref{fig_intro}). This is in contrast with the large majority of segmentation models, which assign one unique label to each voxel \cite{long2015fully, kamnitsas2016efficient}. For OAR segmentation, the previously proposed methods assume either exclusive classes \cite{zhu2019anatomynet,roth2017hierarchical} or non-exclusive classes \cite{nikolov2018deep,wang2018organ,larsson2018robust,ibragimov2017segmentation} similarly to our work. An important difficulty to train machine learning models for multiclass OAR segmentation is the varying availability of ground truth segmentations of different classes among patients, depending on clinical needs. While some organs, such as the optic nerve, are systematically segmented during radiotherapy planning, annotation of other structures may be available only for a subset of patients. One solution to this problem is to independently train one model per class, as it was proposed in some recent deep learning works \cite{larsson2018robust,ibragimov2017segmentation,men2019more}. A limitation of this approach is, however, the need to perform time-consuming trainings for every class, while the number of classes of interest may be large. In this work, we propose a loss function and an algorithm to train neural networks for an end-to-end multiclass segmentation, taking into account the problem of missing annotations. To the best of our knowledge, the only deep learning method for end-to-end multiclass OAR segmentation which addresses this issue is the one proposed in \cite{zhu2019anatomynet} for the segmentation of head and neck organs at risk in CT scans.\newline

The network architecture used in our work is a modified version of 2D U-net \cite{ronneberger2015u}. The choice of a 2D architecture rather than variants of 3D U-Net \cite{cciccek20163d} is motivated by the ability of 2D CNNs to capture a long-range spatial context without downsampling the image. This property is important in our problem as we segment several anatomical structures in large images, including very small structures such as the lens, the pituitary gland or the optic nerve. 2D CNNs were recently applied in \cite{kodym2018segmentation} for segmentation of head and neck organs in CT scans.\newline

Even if most of the proposed deep learning methods for OAR segmentation do not apply anatomical constraints on the output of neural networks, some approaches include shape priors in models. For instance, \cite{tong2018fully} propose to learn latent representations of shapes of organs by a stacked autoencoder and to use these learned representations in the loss function of a segmentation network, in order to compare the shape of the output with the shape of the ground truth. The works \cite{brosch2018deep,orasanu2018organ} propose to adapt triangulated meshes representing organ boundaries to medical images and to use neural networks for regression of distances between centers of triangles and organ boundaries. This type of approach may therefore be seen as atlas-based with the use of deep learning for boundary detection.\newline
However, inclusion of constraints related to connectivity and relative positions of organs in loss functions of CNNs is non trivial due to considerable computational costs. In order to apply such constraints, a neural network would have to segment large regions of the input images during the training phase, which requires a considerable amount of GPU memory. To the best of our knowledge, none of the proposed deep learning methods explicitely enforces consistency of OAR segmentation in terms of relative positions of organs. However, some methods define regions of interest of organs, for instance by registering the image to a set of atlases \cite{larsson2018robust}. \newline

In our work, we enforce some anatomical constraints in a postprocessing stage, starting from the segmentation produced by majority voting of 2D CNNs processing the image by axial, coronal and sagittal slices. In particular, we propose an anatomically consistent segmentation of the optic nerves, with an approach based on the search of the shortest path in a graph, using outputs of neural networks to define weights of edges in the graph.

We consider eight classes of interest, corresponding to anatomical structures systematically segmented during radiotherapy planning for brain cancers: eye, lens, optic nerve, optic chiasm, pituitary gland, hippocampus, brainstem and brain (including cerebrum, cerebellum and brainstem). The anatomical structures composed of left and right components (eye, lens, optic nerve, hippocampus) are seen as one entity by the neural network but are separated in the postprocessing step.

Most of the proposed deep learning methods for segmentation of organs at risk were applied on CT scans in the context of head and neck cancers \cite{argiris2008head}, i.e. cancers of the upper parts of respiratory and digestive systems (mouth, larynx, throat). To the best of our knowledge, the only deep learning method for segmentation of organs at risk in MRIs of the brain is the one proposed in \cite{orasanu2018organ} (MRI T1 and T2).

Our method is tested on a set of contrast-enhanced T1-weighted MRIs acquired in the Centre Antoine Lacassagne in Nice (France). First, our method is quantitatively evaluated on a set of 44 MRIs with provided segmentation of different anatomical structures. Segmentation performances are measured by three different metrics: Dice score, Hausdorff distance and the mean distance between the output and the ground truth. Then, the segmentations produced by our method on a different set of 50 MRIs are qualitatively evaluated by an experienced radiotherapist. Our system was able to produce segmentations with an accuracy level which was found acceptable for radiotherapy planning in a large majority of cases (96\%). The mean distances between the output segmentation and the ground truth for different organs were between 0.1 mm and 0.7 mm.

\section{Methods}
\label{section_methods}
\subsection{Deep learning model}
\subsubsection{Network architecture}
\begin{figure}[t!]
\centering
\includegraphics[width=1.0\textwidth]{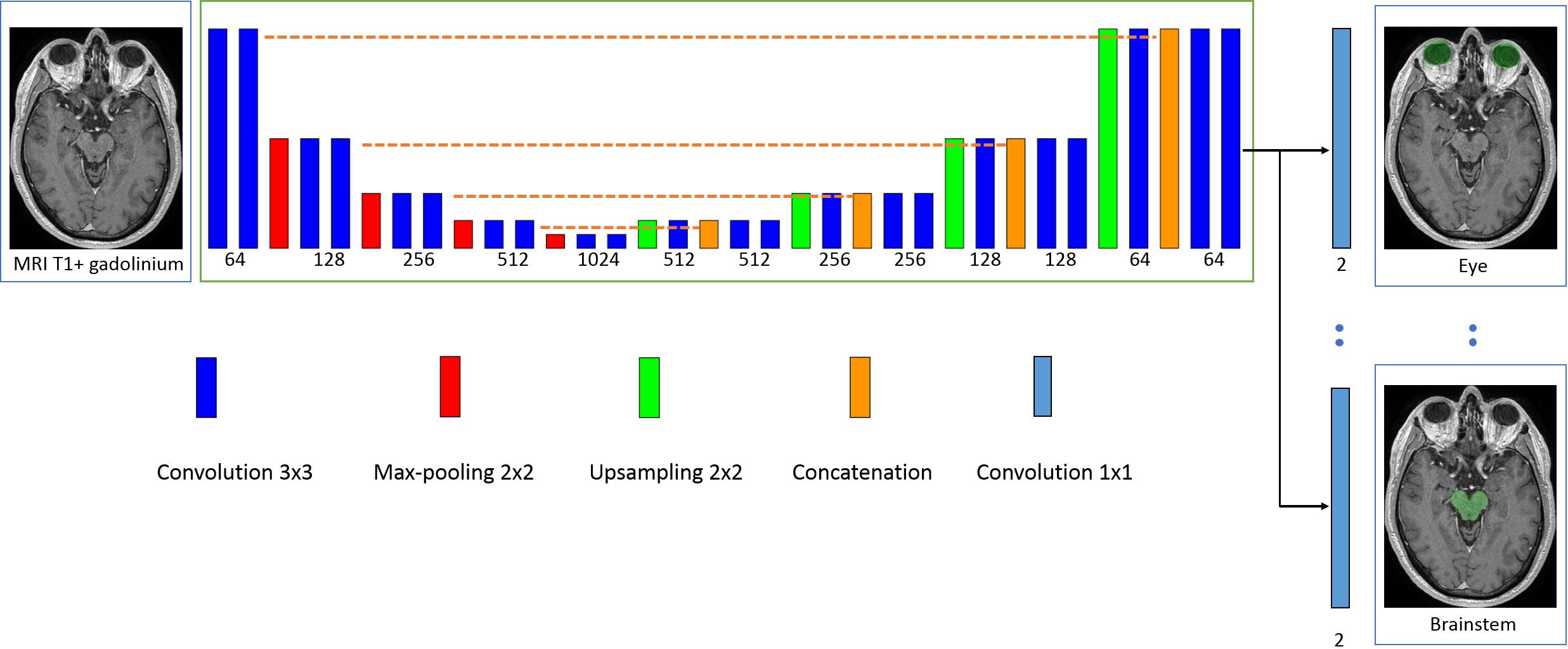}
\caption{Architecture of our model. The rectangles represent layers and their height represents the sampling factor (increasing with max-poolings, decreasing with upsamplings). The numbers of features maps are specified below layers. The proposed model is a modified version of U-Net, having one segmentation layer per class in order to perform an end-to-end multiclass segmentation with non-exclusive classes.}
\label{fig_model}
\end{figure}

The architecture used in our work is a modified version of 2D U-Net \cite{ronneberger2015u}, which is composed of an encoding part and a decoding part. The encoding part is a sequence of convolutional and max-pooling layers. The number of feature maps is doubled after each pooling, taking advantage of their reduced dimensions. The decoding part is composed of convolutional and upsampling layers. Feature maps of the encoding part as concatenated in the decoding part in order to combine low-level and high-level features and to ease the flow of gradients during the optimization process. The final convolutional layer (the segmentation layer) of the standard U-Net has two feature maps, representing pixelwise classification scores of the class 0 ('background') and the class 1. During training, these two final feature maps are normalized by the softmax function.\newline\newline
We adapt this architecture to our problem of multiclass segmentation with non-exclusive classes, where each pixel may belong to zero or several classes. In the following, $C$ denotes the number of classes (in our experiments, $C=8$) and the classes are numbered from 1 to $C$. In our model, each class $c$ has its dedicated  binary segmentation layer (Fig. \ref{fig_model}), composed of two feature maps corresponding to pixelwise scores of the class and of the background. Each segmentation layer takes as input the second to last convolutional layer of U-Net. We use batch normalization \cite{ioffe2015batch} in all convolutional layers of the network, except segmentation layers.

2D CNNs have the advantage of being able to capture information from distant pixels without the need to downsample the input image. In general, CNNs cannot be trained on whole 3D images such as MRI because of their considerable computional costs and GPU memory limitations. In some recent works \cite{roth2017hierarchical,wang2018organ}, 3D CNNs have been applied sequentially in two steps, on a downsampled version of the image and on the original version. As in this work we simultaneously segment several classes which are generally of small size (some may be hardly visible after downsampling) and distant from each other, a 2D architecture is suitable.

\subsubsection{Training of the model}
Our loss function and training scheme were designed to deal with class imbalance and the problem of missing annotations (for a given image, the ground truth is available only for a subset of classes).\newline
In a given training image $i$, each pixel (x,y) has 3 possible labels for the class $c$:  0 (negative), 1 (positive) or -1 (unknown). If the ground truth segmentation of the class $c$ is unavailable for the image $i$, all pixels are labelled as unknown for the class $c$ by default. However, missing annotations may be partially reconstructed from segmentations of other classes. For example, if the segmentation of the 'lens' class is not available but the 'eye' class is segmented, all pixels outside the eye may be labelled as negative for the lens.\newline

Given a training batch of M images and the estimated parameters $\theta$ of the network, the segmentation layer of the class $c$ is penalized by the following loss function, which can be seen as pixelwise cross-entropy with adaptative weights. Let's note $N_0^c$, $N_1^c$ and $N_{-1}^c$ the numbers of pixels labelled respectively 0, 1 and -1 for the class $c$ in the training batch. The weight $w_{(x,y)}^i$ of the pixel (x,y) of the image $i$ has three possible values, according to the label of the pixel. If the label is unknown, then $w_{(x,y)}^i=0$. If its label is 1, then $w_{(x,y)}^i= t_c/ N_1^c $ where $0 < t_c < 1 $ is a fixed hyperparameter, which we call the \textit{target weight}. If the pixel is labelled 0, then $w_{(x,y)}^i= (1-t_c)/ N_0^c $. The introduced hyperparameter $t_c$ controls therefore the relative weight of positive and negative pixels of the class $c$ (positive pixels have the total weight of $t_c$ and negative pixels have the total weight of $1-t_c$). This type of weighting strategy has been used in our previous work \cite{mlynarski20193d} to counter the problem of class imbalance. The loss function of the segmentation layer of the class $c$ is defined by $Loss_{c}(\theta)= - \sum_{ i=1}^{M} \sum_{ (x,y)} w^i_{ (x,y)} \log(p^l_{i, (x,y)}(\theta))$ where $p^l_{i, (x,y)}$ is the softmax score given by the network for the ground truth label $l$ of the pixel.\newline
The loss function of the model is a convex combination of losses of all segmentation layers: $Loss (\theta)= (1/C) \sum_{ c=1}^{C} Loss_{c}(\theta)$.\newline\newline

We propose a sampling strategy to construct training batches so that there are positive and negative pixels for each of the $C$ classes in each training batch.\newline
For each image of the training database, we precompute bounding boxes of all classes with provided segmentations. For bilateral classes such as the eyes, there are generally two bounding boxes per image corresponding to left and right components, unless one of the components is missing (e.g. an organ removed by surgery). The precomputed bounding boxes are used during the training in order to sample patches containing positive pixels of different classes.\newline
At the beginning of the training, for each class $c$, we construct a list $I_c$ of training images with provided ground truth segmentation of the class $c$. To sample a 2D patch which is likely to contain positive pixels of the class $c$, we randomly choose an image $i$ from $I_c$ and a random point $(x,y,z)$ from the bounding box (or two bounding boxes if the class has left and right components) of the class $c$ in the chosen image. Once the point is chosen, a 2D patch (axial, coronal or sagittal) centered on this point is extracted from the image $i$ and segmentations of all available classes are read. In the following, we refer to this procedure as extracting a patch centered on the class $c$. \newline
We assume that the number of images in each training batch ($M$) is larger than the number of classes $C$, in order to be able to sample at least one image/patch centered on each of the classes. Each training batch is constructed as follows. The first $C$ images of the batch are centered respectively on each of the $C$ classes. At this stage, the batch is likely to contain positive and negative pixels of each class. The remaining $M-C$ images may be chosen randomly or be centered on larger classes. In our case, $C=8$, $M=10$ and the last images are centered on the largest class we segment, the brain, whose bounding box occupies almost an entire volume of the head.\newline\newline
As the model is trained for multiclass segmentation with non-exclusive classes, several binary segmentation maps have to be read in each iteration of the training. If the ground truth segmentations are not optimally stored in the memory, these reading operations may considerably slow down the training. The ground truth label of a given pixel can be represented by one bit (0 or 1). However, to store binary segmentation masks in commonly used formats such as HDF5 \cite{folk2011overview}, each label would have to be represented by at least one byte. We propose therefore to store multiclass segmentations in a specifically encoded format, where every bit represents a label of a given class $c$. A binary segmentation mask of the class $c$ is retrieved by the bitwise 'and' operation between the encoded multiclass segmentation and the code of the class, corresponding to a power of 2.\newline
The size of extracted 2D patches should be chosen according to the capacities of the GPU. In our experiments, the training batches were composed of 10 patches of size 230x230. Given that in our network we use unpadded operations (convolutions, max-poolings, etc.), the dimensions of the outputs of segmentation layers are considerably smaller.\newline
The model is trained with a variant of Stochastic Gradient Descent with momentum presented in our previous work \cite{mlynarski20193d}. The main characteristics of this algorithm is that gradients are computed over several batches in each iteration of the training, in order to use many training examples despite GPU memory limitations.

\subsection{Postprocessing and enforcing anatomical consistency}
Fully-convolutional neural networks such as our model produce segmentations by individually classifying every voxel based on intensities of voxels within the corresponding receptive field. Such classification is performed by extracting powerful and automatically learned image features. However, as this classification is performed on a voxel by voxel basis, there is no guarantee of obtaining an anatomically consistent result, especially when the number of training images is limited. In particular, CNNs do not explicitely take into account aspects such as relative positions of different structures or adjacency of voxels belonging to the same structure. Including constraints related to these aspects in loss functions of neural networks or conceiving architectures which produce anatomically consistent results is difficult, in particular because of computational costs (need to simulatenously segment large 3D regions of input images). We propose therefore to improve consistency of segmentations in a postprocessing step. We also separate left and right components of classes such as the eye, as these components are considered separately for radiotherapy planning.\newline
We combine, by majority voting, segmentations produced by three networks trained respectively on axial, coronal and sagittal slices. The goal of this combination is to take into account the three dimensions and to improve the robustness of the method. We subsequently apply a few rules described in the following, in order to correct some observed inconsistencies. \newline

\subsubsection{Segmentation of the brain}
\begin{figure}[t!]
\centering
\includegraphics[width=1.0\textwidth]{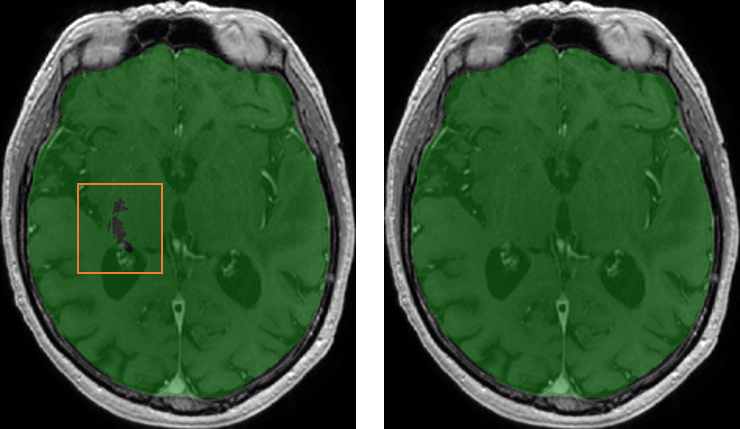}
\caption{Example of 'holes' in the original output segmentation (left image) on a test example. Right image: segmentation obtained after our postprocessing (triplanar hole-filling).}
\label{fig_hole_filling}
\end{figure}

Brain (including the cerebrum, the cerebellum and the brainstem) is the largest class to be segmented. For various reasons, some voxels within this structure may be inconsistently classified as negative by networks,  which appear as 'holes' in the segmentation or unrealistically sharp borders. We propose therefore a procedure that we call triplar hole-filling (Fig. \ref{fig_hole_filling}). For each axial, coronal and sagittal plan of the 3D image, we compute connected components of the background (negative voxels) and we remove components (changing their label from 0 to 1) which are not connected to the border of the plan. The reason of applying this procedure in 2D is that some holes may easily be connected to the outside of the class in 3D. \newline\newline

The bounding box of the segmentation of the brain is subsequently used to separate left and right components of bilateral classes. Please note that the head of the patient may appear at different locations of the image, depending on acquisition conditions and performed preprocessings. For a given class expected to have left and right components (eye, lens, optic nerve, hippocampus), barycenter of each connected component is computed. In order to decide to which side corresponds a connected component, the coordinate x (right-left) of its barycenter is compared to min and max coordinates x of the bounding box of the brain.

\subsubsection{Segmentation of the visual system}
\label{section_postprocessing_visual_system}
We propose an anatomically consistent segmentation of the visual system (eyes, lenses, optic nerves and chiasm), starting from the segmentations predicted by neural networks.\newline
The eye is probably the less challenging organ for automatic segmentation as it has a simple spherical shape. However, some false positives are possible, especially in cases where an eye has been removed by surgery, resulting in false positives within the orbit. We propose therefore to remove connected components of eye segmentation whose volume is below an expected minimum value, which is set to 4 $cm^{3}$.\newline
We constraint segmentation of the lenses to be inside the eyes, i.e. we assign the 0 label to all voxels outside the predicted masks of the eyes.
Segmentation of the optic chiasm is obtained by taking the largest connected component of the segmentation predicted by the networks. We distinguish left and right sides of the chiasm in order to compute landmarks for segmentation of the two optic nerves as described in the following.\newline
Segmentation of the optic nerve in MR images is particularly challenging as the nerve is thin and may have an appearance similar to neighboring structures at some locations. However it has a rather regular shape which can be seen as a tube connecting an eye and the optic chiasm. The nerve is generally well visible at some locations, in particular close to the eye. A human expert is able to track the trajectory of the nerve to distinguish it from neighboring structures at more difficult locations. Based on this observation, we propose a graph-based algorihm for segmentation of the optic nerves in order to guarantee connectivity between the eyes and the optic chiasm and to decrease the number of false positives. The algorithm is based on the search of the shortest path between two nodes in a graph. Outputs of neural networks are used to define weights of the edges in the graph. The different steps of the algorithm (applied separately for left and right nerves) are described below. \newline

\begin{figure}[t!]
\centering
\includegraphics[width=0.7\textwidth]{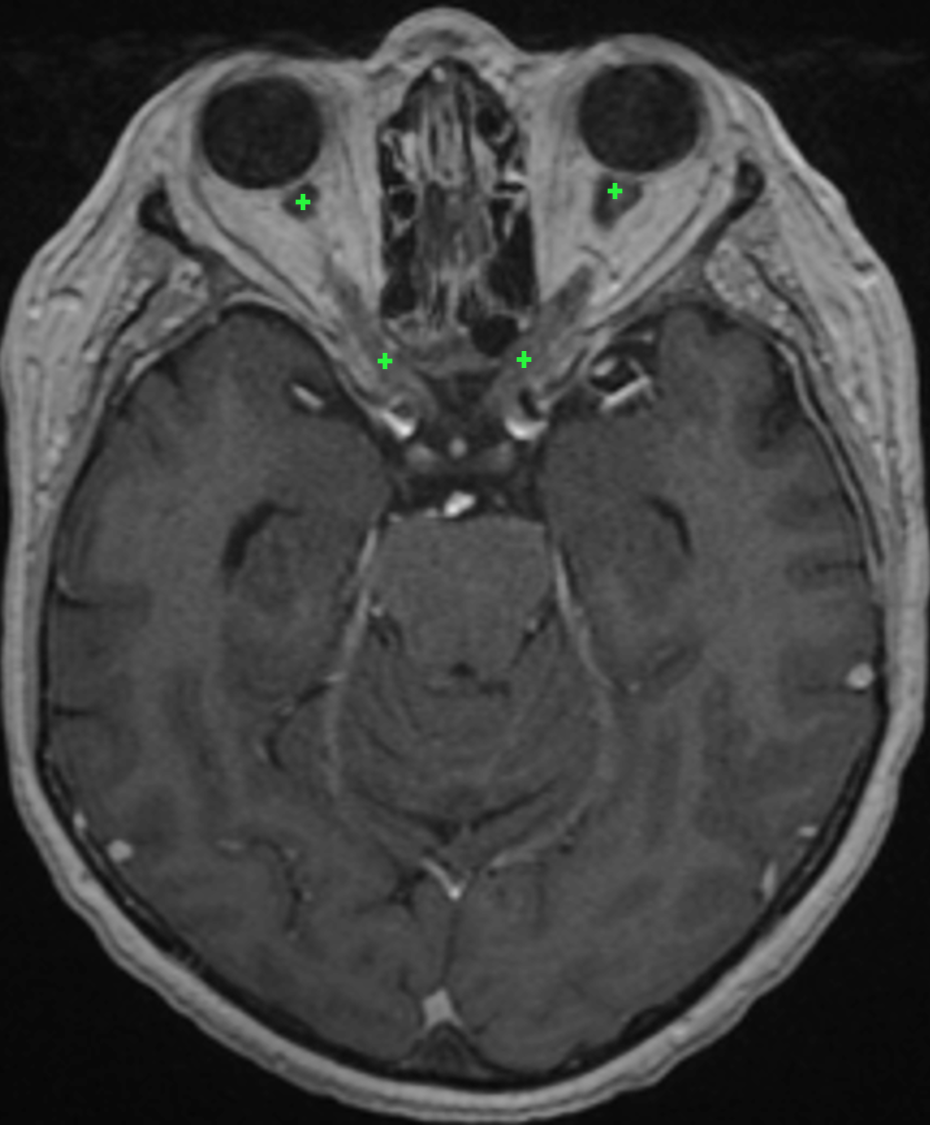}
\caption{Approximate position of the optic nerve landmarks (displayed on the same axial slice) found by the system on a test example. For each of the optic nerves, the graph-based algorithm ensures the connectivity between the two landmarks.}
\label{fig_landmarks}
\end{figure}

First, we detect landmarks corresponding to the two endpoints of an optic nerve based on the initial segmentation of the visual system produced by neural networks (Fig. \ref{fig_landmarks}). The first landmark of the left optic nerve is the barycenter of $P$ points initially predicted as the left optic nerve and which are closest to the left eye. The second landmark is computed similarly but searching $P$ points of the left side of the optic chiasm which are the closest to the initial prediction of the left optic nerve. We take the barycenter of several points  (in our experiments $P=30$) in order to obtain a point which is more likely to be close to the centerline of the nerve. If the detected chiasm landmarks for the two optic nerves are anormally close, the procedure is applied only for one nerve, connecting the landmark with the closest eye.\newline

Before applying the graph-based algorithm, we refine the initial segmentation of the optic nerves based on voxel intensities (specific to each image). In fact, the optic nerves are surrounded by fat, which appears hyperintense on MR T1-weighted images and can be rather easily distinguished from the optic nerve. We compute an approximate range of intensities of voxels of the fat by computing the $98\%$ quantile of a small volume surrounding the eye-nerve landmark. Voxels whose intensites are above $80\%$ of this value are classified negative for the optic nerve in order to eliminate common false positives.\newline

\begin{figure}[b!]
\centering
\includegraphics[width=1.0\textwidth]{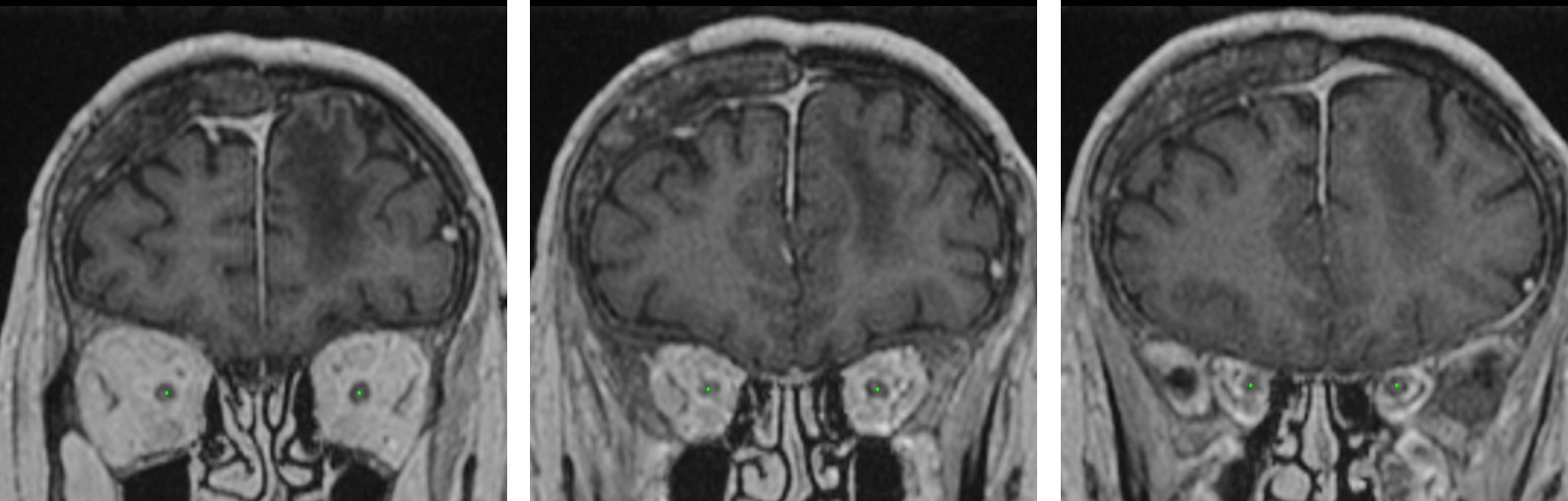}
\caption{The centerlines of the optic nerves computed by our system on a test example (displayed on three different coronal slices). We assume one point of the centerline for each coronal slice between the two landmarks of the optic nerve.}
\label{fig_centerline}
\end{figure}

Given the two computed landmarks and the refined inital segmentation, we estimate the centerline of the optic nerve (Fig. \ref{fig_centerline}) by computing the shortest path in an oriented graph. The nodes of the graph correspond to voxels within a region of interest (cuboid containing the two landmarks) and which are reachable from the starting point. The connectivty of nodes is defined by adjacence of voxels with increasing $y$ coordinate, i.e. the childs of the node $(x,y,z)$ are nodes $(x+d_x,y+1, z+d_z)$ with $d_x \in \{-1,0,1\}$ and $d_z \in \{-1,0,1\}$. We therefore assume strictly increasing $y$ of the centerline towards the second landmark (from anterior to posterior). \newline
Each node $(x,y,z)$ of the graph has its associated cost based on three criteria (listed by decreasing importance):
\begin{itemize}
\item Label $l_{(x,y,z)}$ initially assigned to the voxel $(x,y,z)$. A strong penalty is applied to voxels predicted as negative, in order to force the centerline to pass by points initially predicted as positive. The associated cost is $c^{label}_{(x,y,z)}=0$ if $l_{(x,y,z)}=1$ and $c^{label}_{(x,y,z)}=C^l$ otherwise, where $C^l$ is a fixed number controlling the importance of this cost (we set $C^l=100$).
\item If the predicted label $l_{(x,y,z)}$ is positive: distance $d_{border}$ to the closest point classified as negative. The penalty is inversely proportional to this distance, to give priority to points which are far from predicted borders of the optic nerve (preference to central points). This cost is expressed by $c^{border}_{(x,y,z)}=0$ if $l_{(x,y,z)}=0$ and $c^{border}_{(x,y,z)}=R-d_{border}$ otherwise, where $R$ is the radius of a search zone around the voxel $(x,y,z)$. As the visible nerve is larger close to the eye, $R$ varies with the coordinate $y$ (interpolatation between $R=7$ and $R=3$, expressed in number of voxels). 
\item Distance $d_{target}$ to the target point (i.e. the nerve-chiasm landmark). The penalty is proportional to this distance in order to force the centerline to immediately go towards the target point if other criteria do not give priority to some points. In particular when one part of the optic nerve has not been initially detected (negative voxels), the line should go in the direction of the target point. The associated cost is $c^{distance}_{(x,y,z)}=C^t d_{target}$ where $C^t$ controls the importance of this cost. We fixed $C^l=0.001$, to make it negligible compared to the previous criteria.
\end{itemize}

The cost of the node $(x,y,z)$ is the sum of the three components: {$c_{(x,y,z)}=c^{label}_{(x,y,z)}+c^{border}_{(x,y,z)}+c^{distance}_{(x,y,z)}$}. The introduced cost determines the weights of edges in the graph. A directed edge between the point $(x_1,y_1,z_1)$ and $(x_2,y_2,z_2)$ has the weight of $c_{(x_2,y_2,z_2)}$. The shortest path between nodes corresponding to the two endpoints of the optic nerve is computed by Dijkstra's algorithm \cite{cormen2009introduction,zhan1998shortest}. The start point is the eye-nerve landmark as the optic nerve is generally well visible close to the eye. To the best of our knowledge, our approach is the first to combine deep learning with the search of the shortest path in a graph for segmentation of tubular anatomical structures. However, the idea of computing optimal distances for segmentation of tubular structures appears in interactive level-set methods presented in \cite{deschamps2001fast,cohen1997global,benmansour2011tubular}. The objective of these methods is to find a geodesic between two points in the image chosen by the user. The Eikonal equation is constructed based on voxel intensites and contrasts, and the problem is solved by Fast Marching \cite{sethian1999fast}, similar to Dijkstra's algorithm. Application of methods based only on image intensities may be difficult for segmentation of the optic nerves in MRI due, for instance, to the noise in images and local inhomogeneity of intensities within the optic nerve.

\begin{figure}[t!]
\centering
\includegraphics[width=1.0\textwidth]{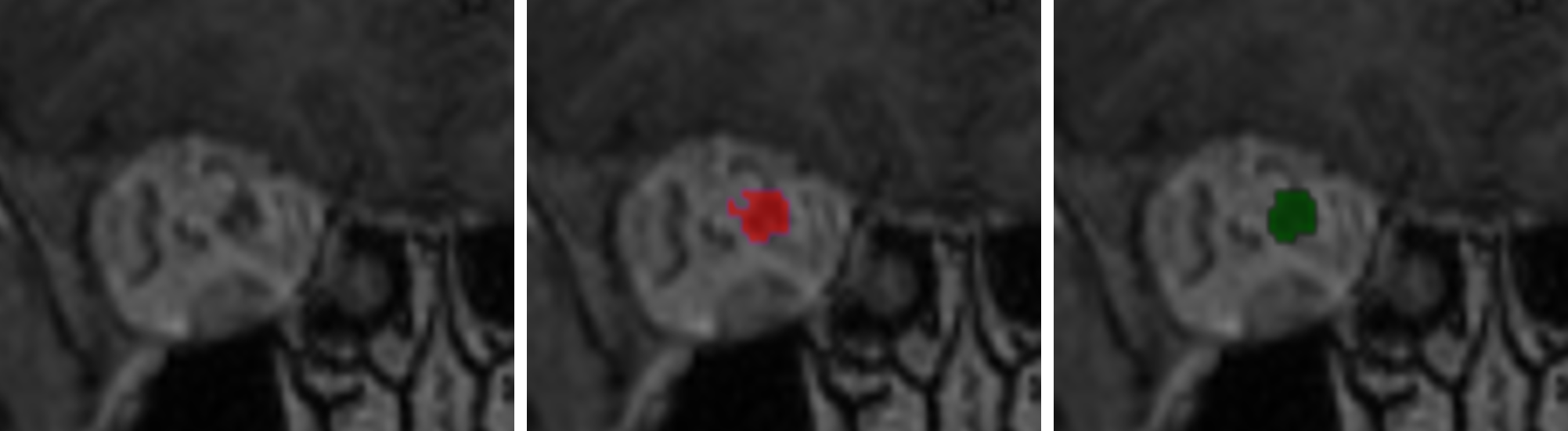}
\caption{Use of mathematical morphology for reduction of false positives. Left: a coronal patch centered on an optic nerve. Middle: result obtained by the system on a test example without using mathematical morphology. Right: result obtained after application of morphological opening followed by taking the largest connected component.}
\label{fig_morphology}
\end{figure}

The final segmentation of the optic nerve is constructed from the centerline. As the optic nerve has a variable thickness, around each point $(x,y,z)$ of the centerline we consider two spherical volumes $S_{(x,y,z)}^1$ and $S_{(x,y,z)}^2$ with associated radii $R_1 \leq R_2$. All voxels within $S_{(x,y,z)}^1$ are classified positive (optic nerve). Voxels of $S_{(x,y,z)}^2$ which are not within $S_{(x,y,z)}^1$ are classified positive only if they were positive in the original segmentation. We fixed $R_1=2.5$ and $R_2$ corresponds to the radius $R$ defined previously (large close to the eye, smaller close to the optic chiasm).\newline
Finally, we apply mathematical morphology \cite{zana2001segmentation} to reduce false positives corresponding to structures which 'attach' to the optic nerve and have a similar appearance. As these false positives are often connected to the correct segmentation by thin segments (Fig. \ref{fig_morphology}), we apply the morphological opening with three 1D structing elements of size 2 in the three directions and we take the largest connected component.

\section{Experiments}
\label{section_experiments}
\subsection{Data and preprocessing}
We constructed a database of contrast-enhanced T1 MRIs acquired in the Centre Antoine Lacassagne (Nice, France), which is one of the three cancer centers in France equipped with proton therapy systems. Proton therapy \cite{levin2005proton} is an external beam radiotherapy which irradiates cancer cells with beams of protons. Ionizing radiation by protons focuses over a narrow range of depth: the energy loss of the radiation follows the \textit{Bragg curve} achieving a pronounced peak just before the particles stop. An important advantage of proton therapy is therefore to offer the possibility to deliver high doses to target volumes while sparing healthy structures. Proton therapy is particularly suitable for treatment of cancers which are very close to critical organs or which are located deep in the body. Currently, one of the main applications of the protontherapy is treatement of brain cancers.\newline
The database contains 44 MRIs with provided segmentations of organs at risk and 50 non-annotated MRIs. The annotated images are used for training and cross-validated quantitative evaluation. For each scan, the ground truth segmentation was provided only for a subset of classes. The numbers of available segmentations for each class are reported in Table \ref{tab_nb_annotations}. The images without annotations are used for qualitative evaluation by a radiotherapist, as described in section \ref{section_clinical_results}.\newline\newline
The images were originally provided in Dicom format \cite{mildenberger2002introduction} and were heterogenous in terms of image intensities and geometrical properties such as size, spatial resolution and the visible part of the head. The ground truth segmentations were the ones used for routine radiotherapy planning and were provided in Dicom RT-Struct files \cite{law2009dicom}, representing coordinates of polygons corresponding to contours of anatomical structures. \newline\newline
In order to use these images, we performed the following preprocessings. We used 3D Slicer \cite{fedorov20123d} and its extension SlicerRT \cite{pinter2012slicerrt} to generate 3D volumes (in \textit{nrrd} format) from Dicom slices and to generate binary label masks from RT-Struct files. All images were resampled to the same spatial resolution, 0.7x0.7x0.9 (isotropic in axial slices, 0.9 spacing between slices) and then resized to dimensions 320x365x200. The 200 axial slices start from the top of the head, i.e. if an image originally has more than 200 axial slices, the bottom slices (close to the neck) are ignored. However, the input images had generally around 150 axial slices and the bottom slices were filled with zeros. To approximately normalize the image intensities, first we compute the maximum of an image, which is likely to be reached by a point on a fat or contrast-enhanced blood vessels. Then, all voxel values are divided by the value of the maximum and multiplied by a fixed constant.

\begin{table}[b!]

\centering
\caption{Numbers of provided ground truth segmentations for different classes (in the database of 44 MRIs).}
\begin{tabular}{|c|c|}
  \hline
&Number of segmentations\\ 
\hline
Hippocampus&39 \\
\hline
Brainstem&39\\  
\hline
Eye&41\\ 
\hline
Lens&34\\ 
\hline
Optic nerves&40\\ 
\hline
Optic chiasm&41 \\ 
\hline
Pituitary gland&29\\ 
\hline
Brain&37\\ 
  \hline
\end{tabular}
\label{tab_nb_annotations}
\end{table}

\subsection{Metrics for quantitative evaluation}
\label{section_metrics}
To quantitatively evaluate our system, we perform 5-fold cross-validation on the set of 44 annotated MRIs. In each fold, 80\% of the database is used for training and 20 \% is used for test. For each class of interest, two results are reported. First, we report results obtained with our model trained on axial slices (denoted 'U-Net multiclass, axial' in the following), i.e. the raw output of the neural network, without postprocessing. Then we report results obtained after majority voting and postprocessing (denoted 'Final result' in the following).\newline\newline
The first metric we use is the Dice score, which measures the voxelwise overlap between the output and the ground truth segmentation.  An important limitation of this metric is that it gives the same importance to very close and very distant mismatches. As the ground truth is often uncertain and noisy close to the boundaries of structures, the Dice scores are generally considerably lower for small structures. This is why, in addition to raw Dice scores, we also report results (Dice, sensitivity, specificity) obtained when a margin of one voxel is allowed, i.e. ignoring mismatches on the borders of the ground truth. This assumption means that a false positive on a voxel $(x,y,z)$ which is directly neighboring with the ground truth segmentation is ignored, i.e. it is neither counted as false positive nor true positive. Similarly, a false negative (non-detection) on the border of the ground truth is ignored.\newline
The second used metric is the undirected Hausdorff distance expressed in millimeters (the coordinates of points are expressed in real values). The Hausdorff distance measures the length of the farthest mismatch between the output and and ground truth (false positive or false negative).  It is therefore useful to assess the consistency of the result, i.e. presence of very distant mistmatches. However, its limitation is that it only measures the value of the maximal distance and therefore one misclassified voxel is sufficient to considerably increase the Hausdorff distance.\newline

Therefore, we also measure the mean distance between the output segmentation A and the ground truth B, defined as follows:
\begin{equation}
M(A,B)= \frac{1}{|A|+|B|}  \left(\sum\limits_{a \in A} \inf\limits_{b \in B} d(a,b) + \sum\limits_{b \in B} \inf\limits_{a \in A} d(b,a)\right)
\end{equation} where $d$ is the Euclidean distance.
\newline\newline

\subsection{Quantitative results}
The mean distances between produced segmentations and the ground truth segmentation ranged from 0.08 mm (for the brain) to 0.69 mm (for the pituitary gland), as reported in Table \ref{tab_results_mean_distance}. The results are variable across the different organs, according to their size, the number of ground truth segmentations available for training and the overall complexity of the segmentation task.\newline
The Dice scores are usually higher for large anatomical structures such as the brain and the brainstem. In particular, the borders of the ground truth are usually very uncertain, which represents a problem for quantitative evaluation for smaller classes. In large classes, the border region is small compared to the entire volume of the class and therefore the mismatches on borders do not cause large drops of the metric. The highest Dice score was obtained for the brain (Dice score of 96.8). The lowest performances were obtained for the pituitary gland (mean Dice of 58, mean distance of 0.69 mm between the output and the ground truth). Segmentation of the pituitary gland is particularly challenging as it is small and difficult to be differentiated from surrounding structures. Moreover, the pituitary gland was the class with the lowest number of training examples (29 annotated cases, i.e. around 23 training cases in each of the 5 folds).\newline
To take into account the uncertain borders of the ground truth, we also reported Dice scores, sensitivity and specificity ignoring mismatches on the border of the ground truth, as described previously. As most mismatches between the outputs and the ground truth are on noisy borders of organs, there is a considerable difference between the raw Dice score (Table \ref{tab_results_dice}) and the Dice score with tolerance to one voxel (Table \ref{tab_results_dice_margin}).\newline
However, the measured Hausdorff distances (Table \ref{tab_results_hausdorff}) are higher for large classes. The highest mean Hausdorff distance is observed for the brain, for which it is of almost equal to 1 cm.

\begin{table}[b!]

\centering
\caption{Mean Dice scores (5-fold cross-validation) obtained on a set of 44 MRIs.}
\begin{tabular}{|c|c|c|}
  \hline
&U-Net multiclass, axial& Final result\\ 
\hline
Hippocampus&69.2& 71.4 \\
\hline
Brainstem&88.1& 88.6\\  
\hline
Eye&88.3&89.6 \\ 
\hline
Lens&55.8&58.8 \\ 
\hline
Optic nerves and chiasm&63.9&67.4 \\ 
\hline
Pituitary gland&53.6&58.0 \\ 
\hline
Brain&96.5&96.8 \\ 
  \hline
\end{tabular}
\label{tab_results_dice}
\end{table}

\begin{table}[b!]

\centering
\caption{Mean Dice score (5-fold cross-validation), sensitivity and specificity with tolerance to one voxel (ignoring mismatches on the borders due to the uncertainty of the ground truth).}
\begin{tabular}{|c|c|c|c|}
  \hline
&Dice score&Sensitivity&Specificity \\ 
\hline
Hippocampus& 88.2&92.7&85.0 \\
\hline
Brainstem& 95.1&95.5&95.6\\  
\hline
Eye&97.5&98.3&96.8 \\ 
\hline
Lens&82.1&88.2&78.4 \\ 
\hline
Optic nerves and chiasm&91.1&96.2&87.1 \\ 
\hline
Pituitary gland&79.7&83.3&77.5 \\ 
\hline
Brain&98.6&98.0&99.4 \\ 
  \hline
\end{tabular}
\label{tab_results_dice_margin}
\end{table}

\begin{table}[b!]

\centering
\caption{Hausdorff distances in millimeters (5-fold cross-validation).}
\begin{tabular}{|c|c|c|}
  \hline
&U-Net multiclass, axial& Final result\\ 
\hline
Hippocampus&42.1&6.9 \\ 
\hline
Brainstem&45.5&7.8 \\ 
\hline
Eye&75.9&3.0 \\ 
\hline
Lens&31.0&3.7 \\ 
\hline
Optic nerves and chiasm&76.7&6.3 \\ 
\hline
Pituitary gland&52.5&4.6 \\ 
\hline
Brain&30.4&9.8 \\ 
  \hline
\end{tabular}
\label{tab_results_hausdorff}
\end{table}

\begin{table}[b!]

\centering
\caption{Mean distances in millimeters  (5-fold cross-validation).}
\begin{tabular}{|c|c|c|}
  \hline
&U-Net multiclass, axial& Final result\\ 
\hline
Hippocampus&0.97&0.66 \\ 
\hline
Brainstem&0.26& 0.26\\ 
\hline
Eye&0.35&0.11 \\ 
\hline
Lens&1.29&0.63 \\ 
\hline
Optic nerves and chiasm&1.09&0.48 \\ 
\hline
Pituitary gland&2.45& 0.69 \\ 
\hline
Brain&0.07&0.08 \\ 
  \hline
\end{tabular}
\label{tab_results_mean_distance}
\end{table}

\begin{figure}[b!]
\centering
\includegraphics[width=0.95\textwidth]{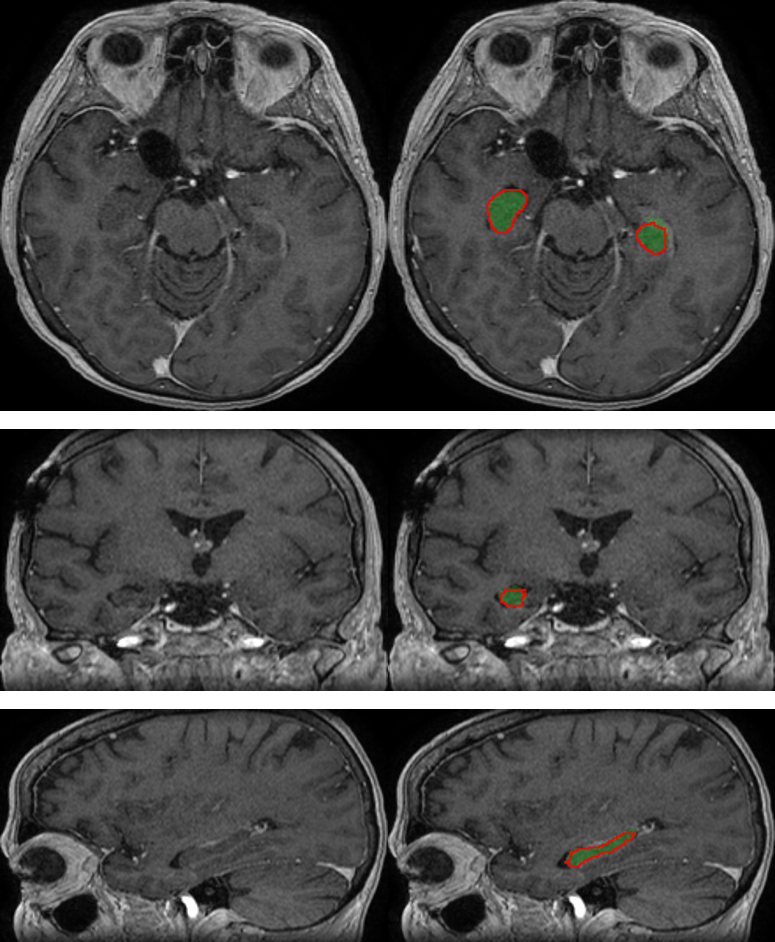}
\caption{Segmentation of the hippocampus produced by our system on a test example (three orthogonal slices passing by the same point). The output segmentation is represented by the green region, the ground truth annotation is represented by the red contour.}
\label{fig_result_hippocampus}
\end{figure}

\begin{figure}[h!]
\centering
\includegraphics[width=1.0\textwidth]{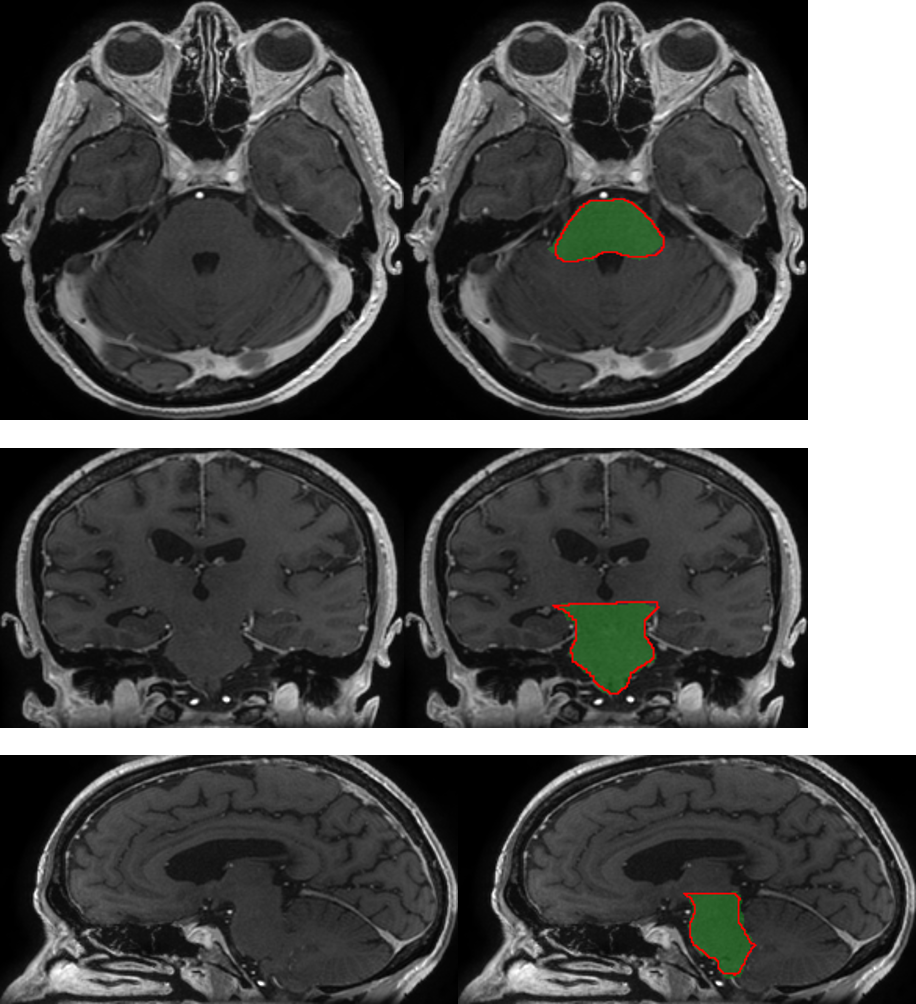}
\caption{Segmentation of the brainstem produced by our system on a test example (three orthogonal slices passing by the same point). The output segmentation is represented by the green region, the ground truth annotation is represented by the red contour.}
\label{fig_result_brainstem}
\end{figure}

The combination of neural networks (trained respectively on axial, coronal, sagittal slices) by majority voting improved almost all metrics. The improvements were particularly large for the Hausdorff distance (Table \ref{tab_results_hausdorff}) and the mean distance (Table \ref{tab_results_mean_distance}). We observe that the majority voting removes almost all distant false positives and yields more robust results than a raw output of one neural network. The results were subsequently improved by additional postprocessings.\newline
The postprocessing of the eyes consisted in setting a lower bound on the physical volume of the output segmentation. This simple procedure allowed to remove false positives and decreased the mean Hausdorff distance from 12.2 mm (result of the majority voting) to 3 mm.\newline
The postprocessing of the optic nerve decreased the number of false positives and enforced connectivity between the eyes and the chiasm, as described in section \ref{section_postprocessing_visual_system}. False positives are removed when they are either too far from the centerline, hyperintense in T1-weighted MRI (fat surrounding eyes) or are disconnected from the main connected component after application of morphological opening removing thin segments. The Dice score with one-voxel tolerance increased from 89.6 (result of the majority voting) to 91.1 (after postprocessing) for the optic nerves and chiasm. The raw Dice score increased from 66.3 to 67.4.\newline
The postprocessing of the brain consisted in taking the largest connected component and filling the 'holes' of the segmentation in axial, coronal and sagittal planes. As these 'holes' are usually small compared to the whole volume of the class (occupying a large part of the image), the variation of the metrics is limited. The Dice score increased from 96.7 to 96.8 and the Hausdorff distance decreased from 10.2 to 9.8.\newline\newline
To the best of our knowledge, the only deep learning work for segmentation of organs at risk in MRI is the one proposed in \cite{orasanu2018organ} which reported cross-validated results (mean distances in mm) on a set of 16 MRIs . The authors used a model-based segmentation \cite{ecabert2008automatic} combined with a neural network for detection of boundaries of anatomical structures. The results reported by the authors for the anatomical structures we also segment are: 0.608 mm for the brainstem, 0.563 mm for the eyes, 0.268 mm for the lenses and 0.41 mm for the optic nerves and chiasm. Overall, the ranges of mean distances are therefore comparable to the ours.\newline\newline
Examples of the output segmentations (comparison to the ground truth) for the hippocampus, the brainstem and the optic nerve are displayed on Fig. \ref{fig_result_hippocampus}, \ref{fig_result_brainstem} and \ref{fig_result_optic_nerve}. The segmentations of others classes are displayed in the supplementary material.

\begin{figure}[h!]
\centering
\includegraphics[width=1.0\textwidth]{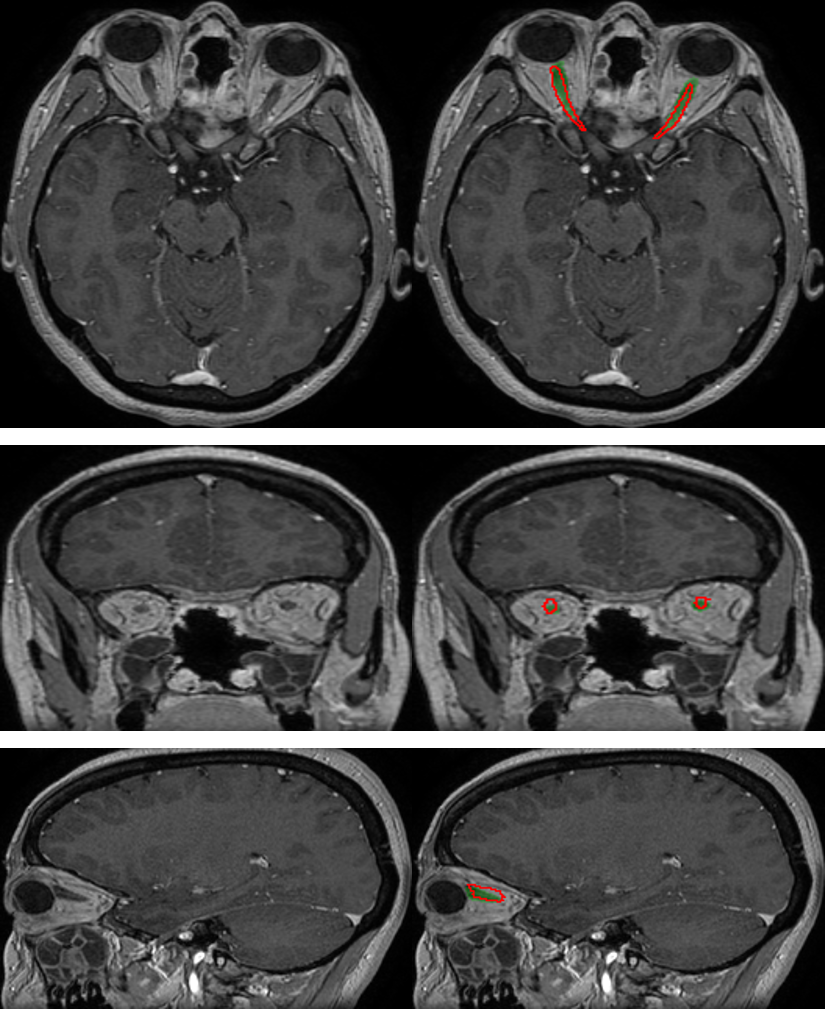}
\caption{Segmentation of the optic nerves produced by our system on a test example (three orthogonal slices passing by the same point). The output segmentation is represented by the green region, the ground truth annotation is represented by the red contour.}
\label{fig_result_optic_nerve}
\end{figure}

\clearpage
\subsection{Qualitative evaluation by a radiotherapist}
\label{section_clinical_results}
The segmentations produced by our system on a set of 50 non-annotated MRIs are qualitatively evaluated by an experienced radiotherapist in order to assess their accuracy and utility for radiotherapy planning. For each of the 50 patients, the radiotherapist qualitatively evaluates the segmentations produced by our system for 12 anatomical structures (counting separately left and right components of bilateral classes), i.e. 600 segmentations are evaluated in total. The segmentations are displayed with 3D Slicer \cite{fedorov20123d}. Each of the 600 segmentations is assigned to one of the following categories:\newline
\begin{itemize}
\item Accept: the radiotherapist would keep the segmentation for radiotherapy planning without any changes
\item Accept with minor modifications: the segmentation is still acceptable for radiotherapy planning, i.e. some minor errors are observed but keeping the current segmentation without changes should not impact the irradiation doses
\item Accept with major modifications: the segmentation has necessarily to be corrected, i.e. keeping the current segmentation would have an important impact on the irradiation doses (even if only few voxels are misclassified). The segmentation is however still good enough to keep it, i.e. it is less time-consuming to perform the necessary modifications than segmenting the structure from the beginning
\item Reject: the segmentation has failed and keeping it would not save time compared to manually segmenting the structure from the beginning
\item Not assigned: the structure is absent (e.g. organ removed by surgery) or invisible in the image because of a tumor
\end{itemize}
The results are summarized in Table \ref{tab_clinical_results}. 73 \% of the segmentations were assigned to the category \textit{accept}, i.e. would be kept for radiotherapy planning without any modifications. Approximately 23 \% of the segmentations were assigned to the second category, i.e. acceptable for radiotherapy planning but with recommendation to perform some minor corrections, usually on extremities of organs. The system produced therefore satisfactory segmentations in a large majority of cases. It was able to correctly delineate organs despite the important difficulties such as presence of  tumors and the resulting mass effects, motion artifacts in MRI, different orientations of heads of patients and anatomical modifications resulting from previous surgeries undergone by the patient (removed tissues).\newline
\begin{table}[t!]

\centering
\caption{Clinical evaluation by a radiotherapist on  50 test cases.}
\begin{tabular}{|c|c|c|c|c|c|}
  \hline
& \footnotesize{Accept}&\footnotesize{Accept, minor}&\footnotesize{Accept, major}&\footnotesize{Reject}&\footnotesize{N/A}\\ 
& & \footnotesize{corrections}& \footnotesize{corrections}&&\\ 
\hline
\small{Hippocampus left}&39/50&8&1&2&0 \\ 
\hline
\small{Hippocampus right}&45/50&5&0&0&0 \\ 
\hline
\small{Brainstem}&22/50&26&1&1&0 \\ 
\hline
\small{Eye left}&48/50&2&0&0&0 \\ 
\hline
\small{Eye right}&45/50&4&1&0&0 \\ 
\hline
\small{Lens left}&39/50&7&4&0&0 \\ 
\hline
\small{Lens right}&42/50&6&2&0&0 \\ 
\hline
\small{Optic nerve left}&44/50&6&0&0&0\\ 
\hline
\small{Optic nerve right}&40/50&10&0&0&0\\ 
\hline
\small{Optic chiasm}&19/50&26&4&1&0\\ 
\hline
\small{Pituitary gland}&19/50&25&3&0&3 \\ 
\hline
\small{Brain}&36/50&14&0&0&0 \\ 
\hline
\hline
\small{Total}&438/600&139&16&4&3 \\ 
  \hline
\end{tabular}
\label{tab_clinical_results}
\end{table}

Segmentations of the eyes had the highest rate of immediate acceptation: 93 out of 100 segmentations were assigned to the \textit{accept} category. The only segmentation which required a major modification was a case with a lesion inside the eye, possibly the polypoidal choroidal vasculopathy. The lesion was not classified by the system as part of the eye and therefore one part of the eye was not detected. The minor modifications recommended for other cases were generally to correct few non-detected voxels on the border of the eye (top or bottom axial slices) or few false positives on the anterior part of the orbit.\newline
All segmentations of the optic nerves were found acceptable for radiotherapy planning: 84 out of 100 segmentations were assigned to the \textit{accept} category and the remaining 16 cases required only minor corrections. Most of the minor errors were non-detections for few voxels on the extremity of the optic nerve close to the eye (e.g. on the top axial slice). There was also at least one case of false positives on the neighboring arteries, close to the optic chiasm.\newline
Even if in the previous, quantitative evaluation, the metrics for the lenses were significantly lower than for other structures, most of their segmentations on the set of 50 MRIs were found satisfactory by the radiotherapist. Minor corrections were required in 13 out of 100 cases and major corrections were required in 6 cases. Most of the problems were non-detections, for instance observed in cases where the patient looks to the side and the system does not detect one side of the lens. The lenses are very small structures and their visibility is highly impacted by motion artifacts in MRI.\newline
For the optic chiasm, the corrections were more frequently required but were usually minor: 19 out of 50 cases were assigned to the \textit{accept} category and 26 cases required minor corrections. The minor errors were often false positives on the hypothalamus (the same issue was observed in the ground truth used to train the model) and sometimes on arteries neighboring the chiasm. The major corrections (4 cases) were mainly non-detections of a small subpart of the beginning of an optic nerve. In fact, even if only a small number of voxels is not detected (false negatives), the corrections are necessary as an excessive irradiation of one part of the optic nerve could make the entire nerve dysfonctional \cite{kallman1992tumour}. One segmentation was rejected due to non-detection of one part of the chiasm. This error appeared in a challenging case where the anatomy of the patient was modified by an important mass effect caused by a tumor.\newline
Similar performances were obtained for the pituitary gland, located below the optic chiasm. Most of the minor (26 cases) and major (3 cases) required corrections correspond to non detections, typically on the 1-2 lowermost slices. In at least 2 cases, few false negatives were observed on the pituitary stalk (also observed in some ground truth segmentations used for training), which is the connection between the pituitary gland and the hypothalamus.\newline
Even if segmentation of the hippocampus is difficult (low contrast with neighboring structures), in our evaluation it had one of the highest acceptation rates, with 84 segmentations in the \textit{accept} category. However, it is also the only structure for which more than one segmentation was rejected. The two rejected segmentations correspond to cases where a large tumoral mass has grown near to the hippocampus, causing an edema having a similar intensity in T1-weighted MRI. Morever, the tumors had a large necrotic core which may be confused with a ventricle by the system. In other cases, the required corrections (mostly minor) correspond usually to false positives (in particular on the amygdales, neighboring hippocampi and having a similar intensity in MRI T1) or some non-detections on the extremities of the hippocampus.\newline
For the brainstem, 48 out of 50 segmentations were found acceptable for radiotherapy planning but required minor modifications in approximately half cases. The required corrections (false positives or non-detections) were almost exclusively on the uppermost axial slices (typically on 2 slices) which correspond to the top extremity of the brainstem. The only rejected segmentation corresponds to a case with a tumor adjacent to the brainstem and which was mistakenly included in the segmentation (false positives).\newline
Finally, all segmentations of the brain (occupying a large part of the head) were found acceptable for radiotherapy planning even if they required minor corrections in almost one third of cases. The recommended corrections include, for instance, non-detections close to the cribriform plate (between the eyes) and false positives on bones.\newline

In particular, we observe that the only two structures for which all segmentations were found acceptable for radiotherapy planning (without any major correction) are the ones for which a specific postprocessing was performed, i.e. the optic nerves and the brain.

\section{Conclusion and future work}
\label{section_conclusion}
In this paper we proposed a CNN-based method for segmentation of organs at risk from MR images in the context of neuro-oncology. The method was evaluated on clinical data.\newline
First, we proposed a deep learning model and a training algorithm for segmentation of multiple and non-exclusive anatomical structures. The proposed methodology addresses problems related to computational costs and the variable availability of ground truth segmentations of the different anatomical structures (unsegmented classes). The neural network used in our method is a modified version of U-Net. The network is trained separately for segmentation in axial, coronal and sagittal slices. The three versions of the network are combined by majority voting.\newline
Second, we proposed procedures to enforce anatomical consistency of the result in a postprocessing stage. In particular, we proposed a graph-based algorithm for segmentation of the optic nerves, which are among the most difficult anatomical structures for automatic segmentation. The proposed postprocessings have shown their efficiency particularly in the qualitative evaluation by a radiotherapist. In particular, all segmentations of the optic nerves were found acceptable for radiotherapy planning.\newline
The method was evaluated quantitatively on a set of 44 annotated MRIs, with 5 fold cross-validation and using several metrics. The segmentations produced by our system on a set of 50 non-annotated MRIs were qualitatively evaluated by an experienced radiotherapist. Despite the limited size of the training database (44 annotated MRIs) and the different challenges of the segmentation tasks (in particular, presence of tumors), a large majority of the output segmentations were found sufficiently accurate to be used for computation of irradiation doses in radiotherapy.\newline\newline
An important step of the future work is to adapt the method to multimodal data. Often, several types of images are acquired during radiotherapy planning for one patient, including the different MR sequences (T1, T2, FLAIR) and CT scans. Inclusion of different imaging modalities could improve segmentation of several structures but it comes also with new challenges related, for instance, to inter-modality registration and training of models on cases with missing modalities.\newline
As for other segmentation tasks in medical imaging, availability of annotated training data is an important problem. Methods able to exploit weaker forms of annotations (bounding boxes, slice-level labels) for training of segmentation models are therefore of interest. In particular, methods combining weakly-annotated and fully-annotated training images were recently proposed in \cite{mlynarski2018deep,shah2018ms}. As our system was able to produce accurate segmentations in a large majority of cases and the rare observed errors were mainly on boundaries of organs, the system could be used for generation of bounding boxes (subsequently verified by a human) which could be used to train segmentation models which are able to exploit this type of annotations.\newline
Another important direction of the future work is to combine segmentation of organs at risk and segmentation of radiotherapy target volumes. In particular, a large variability of methods for tumor segmentation \cite{myronenko20183d,kamnitsas2017efficient,wang2017automatic,mlynarski20193d,parisot2014concurrent} were proposed in recent years. Deep learning could also be used for computation of irradiation doses \cite{andres2019po} in radiotherapy planning.

\section*{Acknowledgements}
Pawel Mlynarski is funded by the Microsoft Research-INRIA Joint Centre, France. This work was supported by the Inria Sophia Antipolis - M\'editerran\'ee, "NEF" computation cluster.


\clearpage
\section*{Supplementary material}
\begin{figure}[b!]
\centering
\includegraphics[width=0.87\textwidth]{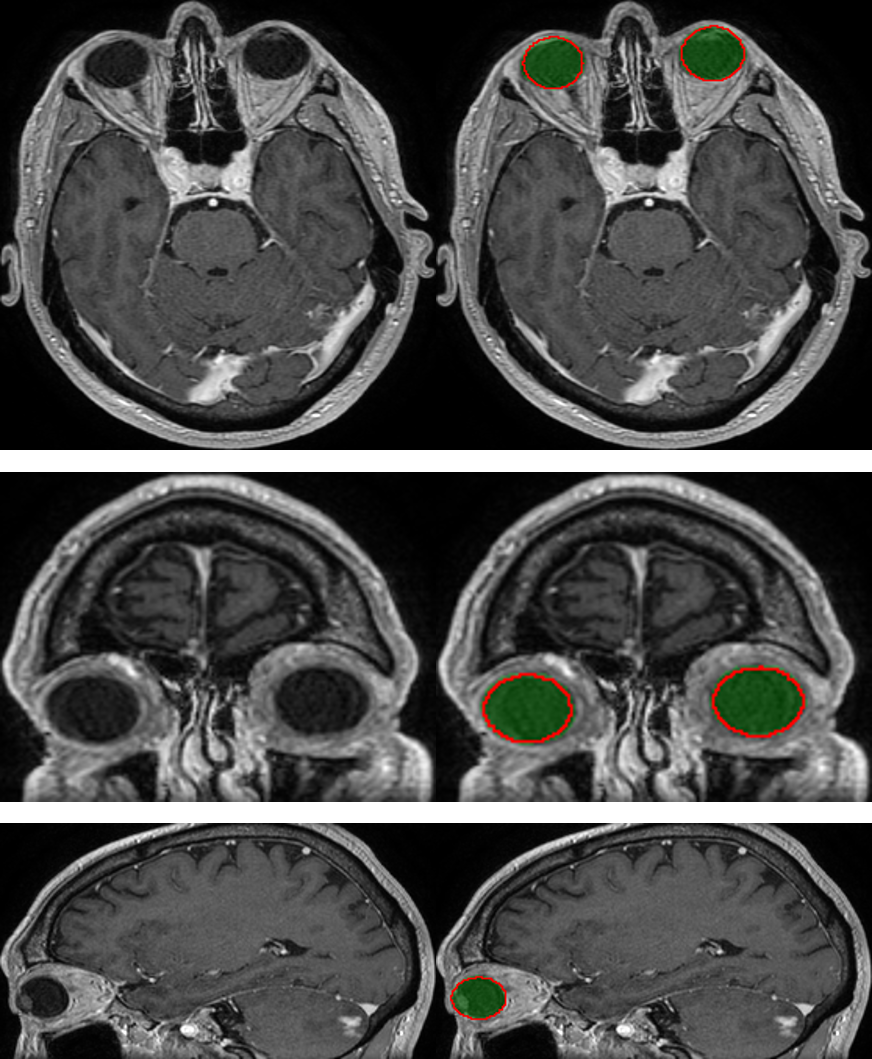}
\caption{Segmentation of the eyes produced by our system on a test example (three orthogonal slices passing by the same point). The output segmentation is represented by the green region, the ground truth annotation is represented by the red contour.}
\label{fig_result_eye}
\end{figure}

\begin{figure}[t!]
\centering
\includegraphics[width=1.0\textwidth]{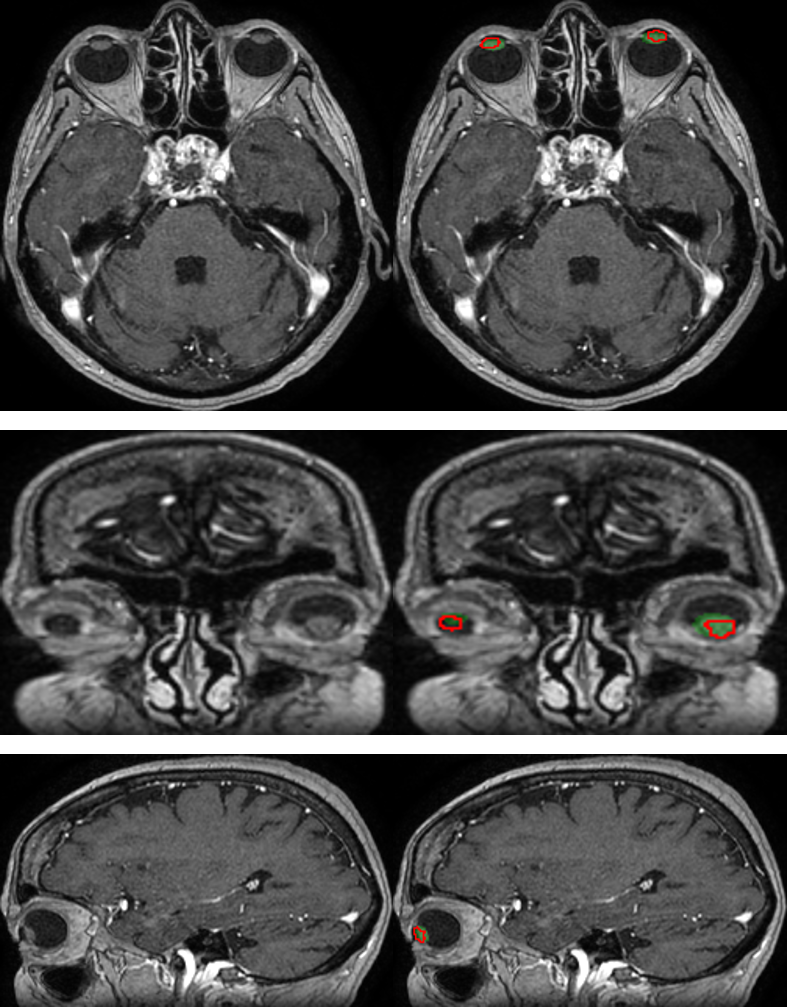}
\caption{Segmentation of the lenses produced by our system on a test example (three orthogonal slices passing by the same point). The output segmentation is represented by the green region, the ground truth annotation is represented by the red contour.}
\label{fig_result_lens}
\end{figure}

\begin{figure}[h!]
\centering
\includegraphics[width=1.0\textwidth]{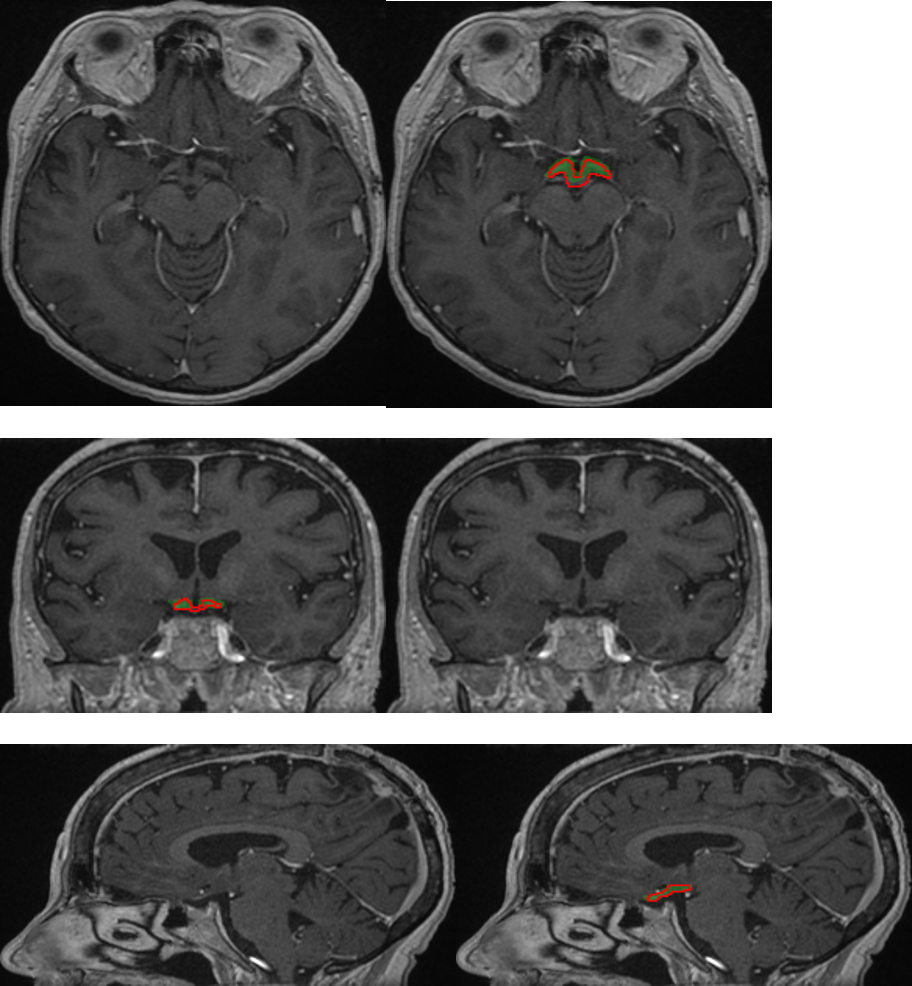}
\caption{Segmentation of the optic chiasm produced by our system on a test example (three orthogonal slices passing by the same point). The output segmentation is represented by the green region, the ground truth annotation is represented by the red contour.}
\label{fig_result_chiasm}
\end{figure}

\begin{figure}[h!]
\centering
\includegraphics[width=1.0\textwidth]{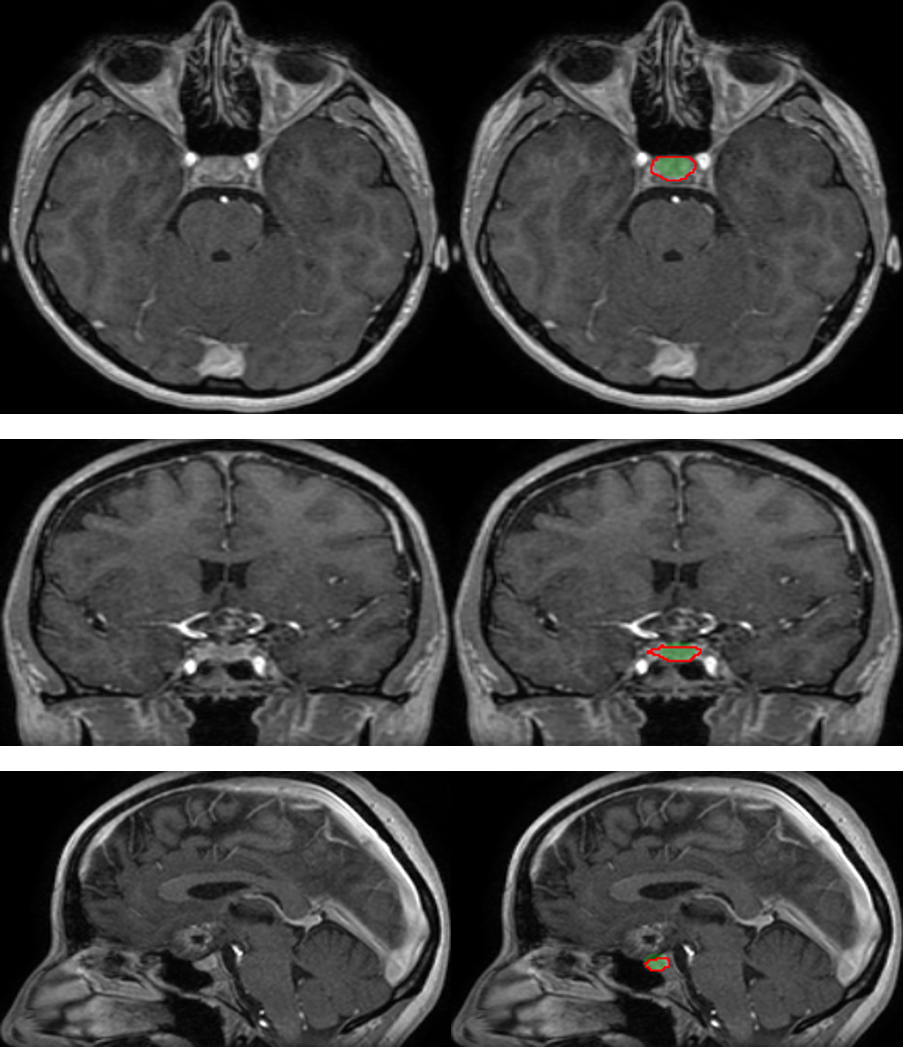}
\caption{Segmentation of the pituitary gland produced by our system on a test example (three orthogonal slices passing by the same point). The output segmentation is represented by the green region, the ground truth annotation is represented by the red contour.}
\label{fig_result_hypophysis}
\end{figure}

\begin{figure}[h!]
\centering
\includegraphics[width=1.0\textwidth]{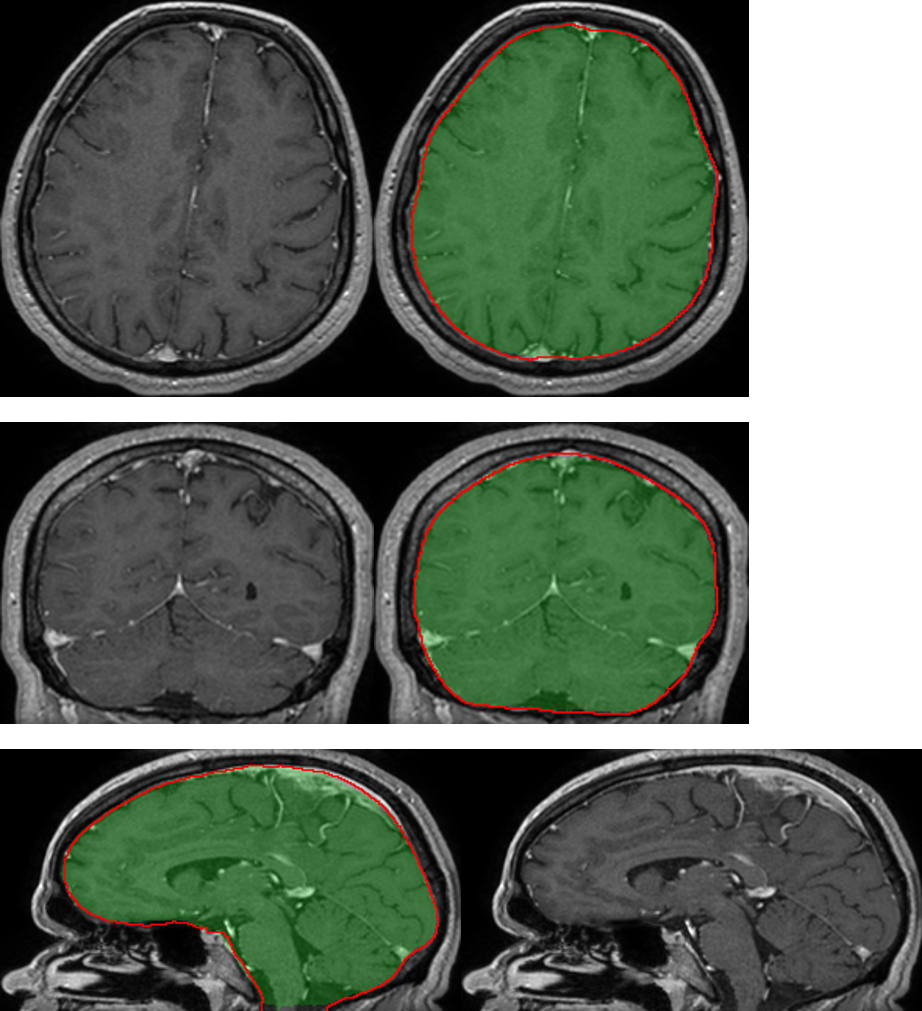}
\caption{Segmentation of the brain produced by our system on a test example (three orthogonal slices passing by the same point). The output segmentation is represented by the green region, the ground truth annotation is represented by the red contour.}
\label{fig_result_brain}
\end{figure}

In this supplementary material, we display examples of the output segmentations for the following classes: eye (Fig. \ref{fig_result_eye}), lens (Fig. \ref{fig_result_lens}), optic chiasm (Fig. \ref{fig_result_chiasm}), pituitary gland (Fig. \ref{fig_result_hypophysis}) and brain (Fig. \ref{fig_result_brain}).


\clearpage

\bibliographystyle{apalike} 
\bibliography{biblio}

\end{document}